\documentclass{article}

\usepackage[T1]{fontenc}
\usepackage[margin=1in]{geometry} 
\usepackage{graphicx,verbatim}
\usepackage{cite}
\usepackage{amsmath}
\usepackage{amssymb}
\usepackage{subcaption}
\usepackage{float} 
\usepackage{xcolor}

\title{Evaluating structural uncertainty in accelerated MRI: are voxelwise measures useful surrogates?}

\author{
Luca L. C. Trautmann$^{1}$\thanks{ORCID: 0009-0001-3517-5471} \and
Peter A. Wijeratne$^{1}$\thanks{ORCID: 0000-0002-4885-6241} \and
Itamar Ronen$^{2}$\thanks{ORCID: 0000-0002-6872-4895} \and
Ivor J. A. Simpson$^{1}$\thanks{ORCID: 0000-0001-5605-6626}
}

\date{
$^{1}$ SussexAI Centre, Department of Engineering and Informatics, Brighton, UK\\
$^{2}$ Brighton and Sussex Medical School, Brighton, UK
}

\begin{document}
\maketitle
\begin{abstract}
Introducing accelerated reconstruction algorithms into clinical settings requires measures of uncertainty quantification that accurately assess the relevant uncertainty introduced by the reconstruction algorithm. Many currently deployed approaches quantifying uncertainty by focusing on measuring the variability in voxelwise intensity variation. Although these provide interpretable maps, they lack a structural interpretation and do not show a clear relationship to how the data will be analysed subsequently. In this work we show that voxel level uncertainty does not provide insight into morphological uncertainty. To do so, we use segmentation as a clinically-relevant downstream task and deploy ensembles of reconstruction modes to measure uncertainty in the reconstructions. We show that variability and bias in the morphological structures are present and within-ensemble variability cannot be predicted well with uncertainty measured only by voxel intensity variations.

\end{abstract}
\section{Introduction}
Magnetic Resonance Imaging (MRI) is a key medical imaging technique, offering non-invasive, non-ionising visualization of tissue for diagnosis, surgical planning, and research. However, MRI faces a fundamental constraint: improving image quality generally requires longer acquisition times, which increase patient discomfort, motion artefacts, and limit scanner throughput. To address these limitations, accelerating MRI acquisition through partial sampling of measurements (k-space) is an active area of research \cite{hammernik_physics-driven_2022, chung_score-based_2022, heckel_deep_2024, sriram_end--end_2020}. 

Early work on signal recovery from undersampled k-space focused on compressed sensing (CS) and parallel imaging (PI) to reduce artefacts. However, both approaches impose additional requirements, such as specific undersampling patterns, that are often impractical in clinical settings \cite{hammernik_learning_2017, heckel_deep_2024}. More recently, deep learning (DL)-based reconstruction methods have shown promise in overcoming these trade-offs by producing high-fidelity, detail-rich images from undersampled data through data-driven learning. These DL approaches offer the potential to significantly reduce scan times, lower costs, and improve MRI accessibility, while mitigating existing incompatibilities as present in CS and PI. 

Regardless of the applied reconstruction method, image reconstruction from undersampled k-space, is necessarily ill-posed, as missing data needs to be inferred \cite{cohen_looks_2024}. For real-world applications, this inherent ill-posedness of the reconstruction problem, therefore, demands robust, clinically relevant uncertainty quantification before it can be reliably introduced into practical deployment. Currently, where uncertainty estimates are provided, they are typically expressed as pixel-wise or voxelwise maps that represent intensity-based uncertainty variations. In this paper, we argue that treating uncertainty or error in an independent voxelwise manner is poorly aligned with clinically relevant volume uncertainty of structural MRI. Our central claim is that clinically meaningful uncertainty quantification requires task-aware metrics and tools aligned with the diagnostic goals that motivate imaging in the first place. We show that intensity-based uncertainty estimates fail to capture the bias, magnitude, and spatial distribution of morphological variation in downstream tasks. Focusing on segmentation as a representative task, we analyse sample reconstructions drawn from an ensemble of model outputs. We use SynthSeg \cite{billot_robust_2023} as a deterministic, one-shot segmentation tool, reflecting typical clinical use.

\subsection{Contributions}
In this work, we make the following contributions:
\begin{enumerate}
\item We demonstrate that segmentations of ensemble-based reconstruction models exhibit bias and variability, even when individual models achieve high SSIM or PSNR scores.
\item We show, through correlation and regression analysis, that voxel intensity uncertainty does not predict volume uncertainty within ensembles in the downstream segmentation task, nor bias and uncertainty relative to the fully sampled segmentation.
\end{enumerate}

\section{Background}
\subsection{Problem specification}
Consider a typical linear inverse problem \cite{hammernik_systematic_2021, arridge_solving_2019, arridge_inverse_2024} 
\begin{equation}
    \mathbf{y} = \mathbf{A} \mathbf{f} + \mathbf{e}, \quad 
    \mathbf{y} \in \mathbb{C}^m, 
    \mathbf{A} \in \mathbb{C}^{m \times n},  
    \mathbf{f} \in \mathbb{C}^n, 
    \mathbf{e} \in \mathbb{C}^m.
    \label{eq:inverse_problem}
\end{equation}

where true signal $\mathbf{f}$ is modified by a known forward model $\mathbf{A}$ to produce the measurements $\mathbf{y}$. $\mathbf{e}$ represents additional noise added in the transformation. The reconstruction algorithm must find an appropriate inversion of the forward model \(\mathbf{A}\) to recover the true signal from the measurements. In the case of accelerated MRI, $\mathbf{y}$ are k-space measurement and $\mathbf{f}$ represents the true image at acquisition time. The undersampling of the k-space leads to equation (\ref{eq:inverse_problem}) becoming underdetermined and generally ill posed, whereby no clear one-to-one mapping between the undersampled measurements and the true signal exists\cite{hammernik_physics-driven_2022}.
 
\subsection{Uncertainty in ill-posed reconstruction problems}
\begin{figure}
    \centering
    \includegraphics[width=0.75\linewidth]{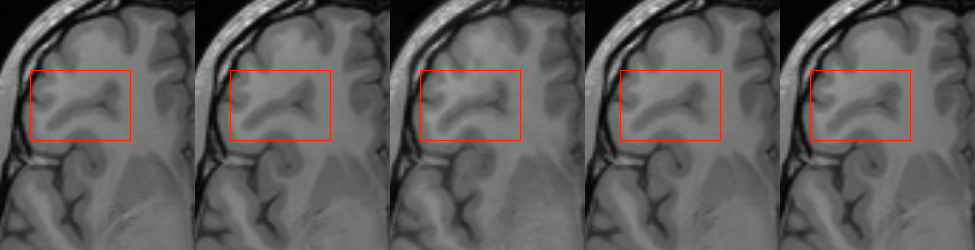}
    \caption{Example slices from 3D reconstruction volumes. Within each red bounding box subtle differences produced by differently initialised Variational Network (VN) ensemble members from identically undersampled k-space are visible. Variations include white and grey matter thickness and shifted boundaries.}
    \label{fig:Reconstruction_Differences}
\end{figure}
Uncertainty in deep learning can be divided into irreducible data (aleatoric) and reducible epistemic (model) uncertainty \cite{abdar_review_2021}. Two main approaches capture this uncertainty: Bayesian neural networks (BNNs) and ensembles \cite{gawlikowski_survey_2023}. BNNs place priors over parameters and perform Bayesian inference to obtain a posterior predictive distribution encoding both epistemic and aleatoric uncertainty \cite{loaiza-ganem_deep_2025}. Ensemble methods approximate this distribution by averaging predictions from independently trained models with different initialisations \cite{ganaie_ensemble_2022}. While BNNs are often seen as more principled due to their formal Bayesian grounding, research has established rigorous connections between the two approaches \cite{hoffmann_deep_2021, wild_rigorous_2023, loaiza-ganem_deep_2025}. In this work, we use deep ensembles with existing reconstruction models to examine uncertainty introduced during reconstruction from undersampled k-space.

As mentioned before, any deep-learning-based reconstruction of undersampled MRI data inherently carries uncertainty, because it must infer missing information \cite{cohen_looks_2024} in addition to handling measurement noise. Model predictions are influenced by inductive biases built into the architecture \cite{wilson_bayesian_2022}, as well as by parameter initialisation and training hyperparameters \cite{prince_understanding_2023}. Figure \ref{fig:Reconstruction_Differences} illustrates subtle differences between different models in an ensembled Variational Network (VN) trained with different random initialisation. Quantifying epistemic uncertainty aims to measure the effects of these factors on model outputs. Adjacent to epistemic uncertainty is prediction bias. Bias is especially relevant in structural imaging, where the goal is typically diagnosis or planning for intervention. Both depend on accurate representations of the shape and location of structures.

Most existing methods assess uncertainty at the pixel or voxel level, while single image quality is typically evaluated using SSIM or PSNR. Uncertainty maps usually show voxelwise intensity variations \cite{narnhofer_bayesian_2021, chung_score-based_2022, edupuganti_uncertainty_2020, kustner_predictive_2024}. However, in structural MRI, such intensity-based metrics are not always well-suited. MRI intensities are not inherently quantitative, making variations difficult to meaningfully interpret \cite{wen_task-driven_2024}. Voxel-level variances also lack semantic value, are highly sensitive to noise and scaling, and assume independence between voxels, ignoring anatomical structure and spatial coherence. Despite these limitations, they remain the dominant tools for assessing reconstruction quality and epistemic uncertainty. 

\section{Methodology}
\subsection{Data} 
We used the Calgary Campinas 2022 challenge dataset \cite{souza_open_2018}. This multi-channel 3D T1-weighted MRI dataset includes 167 scans from presumed healthy subjects (age: $44.5 \pm 15.5$ years), acquired on a GE Discovery MR750 scanner. For the reconstruction model, we use only the 12-channel (117 scans) coil at 1 mm isotropic resolution ($256 \times 218 \times [170, 180]$ matrix). Acquisition parameters were $TR / TE / TI = 6.3 / 2.6 / 650\, \mathrm{ms}$ (93 scans) and $7.4 / 3.1 / 400\, \mathrm{ms}$ (74 scans), with partial Fourier sampling ($85\%$) in the slice direction. The dataset is split according to the original challenge design into training (47), validation (20). All experiments were conducted on the validation dataset. Bilateral structures in the segmentation output were combined by summing left and right hemisphere volumes.

\subsection{Models}
We ensembled three different model architectures from the 2022 Calgary Campinas MRI reconstruction challenge with DIRECT \cite{yiasemis_direct_2022}. We selected models along the performance curve with different intrinsic biases: a $\text{UNET}_\text{k-space}$ without data consistency, Variational Networks (VN) \cite{kendall_what_2017}, and Recurrent Inference Machines (RIM) \cite{putzky_recurrent_2017, lonning_recurrent_2019}. Within each ensemble, models were trained with identical hyperparameters but different random initialisations (see Appendix for training specifics). The UNet ensemble included 15 runs, with mean PSNR 30.65 ($\pm$ 0.18) and SSIM 0.8877 ($\pm$ 0.0024). VN used 11 runs, achieving PSNR 33.40 ($\pm$ 1.09) and SSIM 0.9164 ($\pm$ 0.0127). The RIM ensemble comprised 11 runs with PSNR 35.05 ($\pm$ 0.04) and SSIM 0.9329 ($\pm$ 0.0003). All data was 5x accelerated, and both input and output resolutions were isotropic with 1mm voxel size. SynthSeg \cite{billot_robust_2023} was run in robust mode, and we used the volume estimates provided by SynthSeg.
 
\section{Experiments and Results}
We assessed the relationship between voxel-level uncertainty and morphological uncertainty in segmentations, focusing on structures with clinically relevant volume measures (a subset of the 33 SynthSeg labels). All analyses were run on the validation set, with results aggregated across all subjects.

\subsection{Volume Variation and Bias Analysis}

\begin{figure}[h!]
    \centering
    \includegraphics[width=0.75\linewidth]{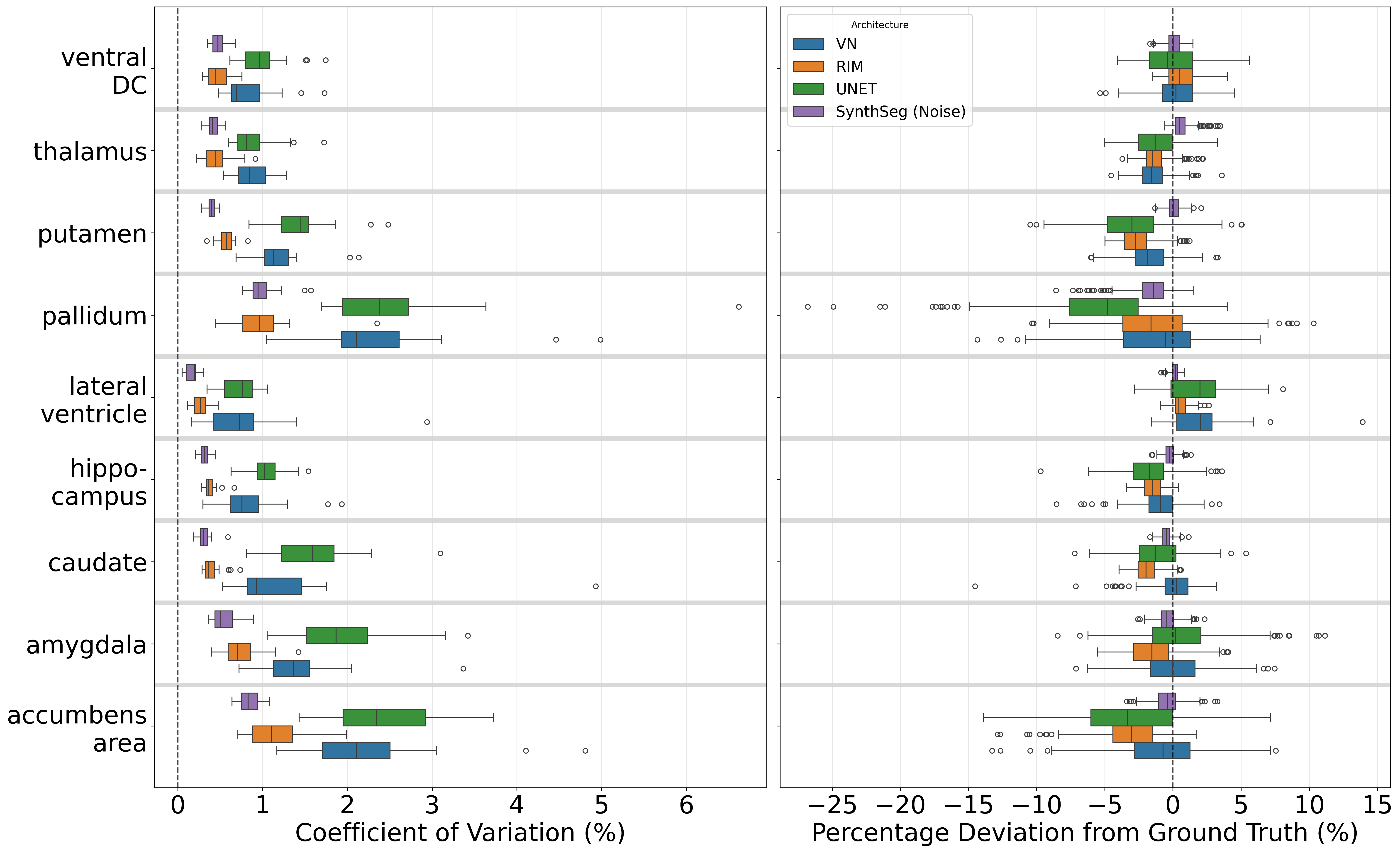}
    \caption{\textbf{Left:} shows ensemble uncertainty, quantified as the coefficient of variation of predicted structure volumes across 20 validation subjects. Box plots display the distribution of subject-level CV values for each structure and reconstruction architecture, with bilateral structures summed. Lower CV indicates greater internal consistency among ensemble members. SynthSeg (Noise) indicates the CV for a fixed SNR of 30 of fully sampled measurements due to noise. \textbf{Right:} shows prediction bias relative to fully-sampled ground truth, expressed as percentage deviation. Box plots show the distribution of deviations across subjects and all predictions in the ensemble, revealing systematic bias (median deviation from zero) and variability.}
    \label{fig:Distribution of CV values across samples}
\end{figure}

To quantify ensemble uncertainty in predicted brain structure volumes, we measured the coefficient of variation (CV) of volumes across ensemble predictions for each structure. The CV, defined as the ratio of standard deviation to mean, provides a normalized measure of variability independent of structure size. As a baseline, we created synthetic ensembles by adding Gaussian noise (SNR=30) to fully sampled k-space data to evaluate variability due to measurement noise alone. Figure \ref{fig:Distribution of CV values across samples} (left) shows the distribution of CV values across subjects for different architectures. All models exhibited within-ensemble volume variation in clinically relevant structures, with RIM showing the lowest and UNET the highest variation. Across models, volume differences generally ranged from 0.5\% to 3.5\%, with UNET and Variational Network (VN) showing occasional higher outliers. Noise-induced variation (SynthSeg (Noise)) was consistently lower than the variation observed in learned reconstructions, indicating that model-induced uncertainty dominates measurement noise.

To assess systematic errors, we measured bias as the percentage volume difference between segmentations of reconstructed images and fully sampled references. Figure \ref{fig:Distribution of CV values across samples} (right) displays the distribution of these biases across subjects. While RIM exhibited lower variance, it showed greater bias, consistently underestimating structures such as the thalamus, putamen, hippocampus, and caudate compared to the ground truth. Notably, RIM, despite consistent volume variance, showed greater bias, underestimating the thalamus, putamen, hippocampus, and caudate volumes compared to fully sampled segmentations.

\subsection{Voxelwise Uncertainty and Volume Variability}

\begin{figure}[h!]
    \centering
    \includegraphics[width=0.75\linewidth]{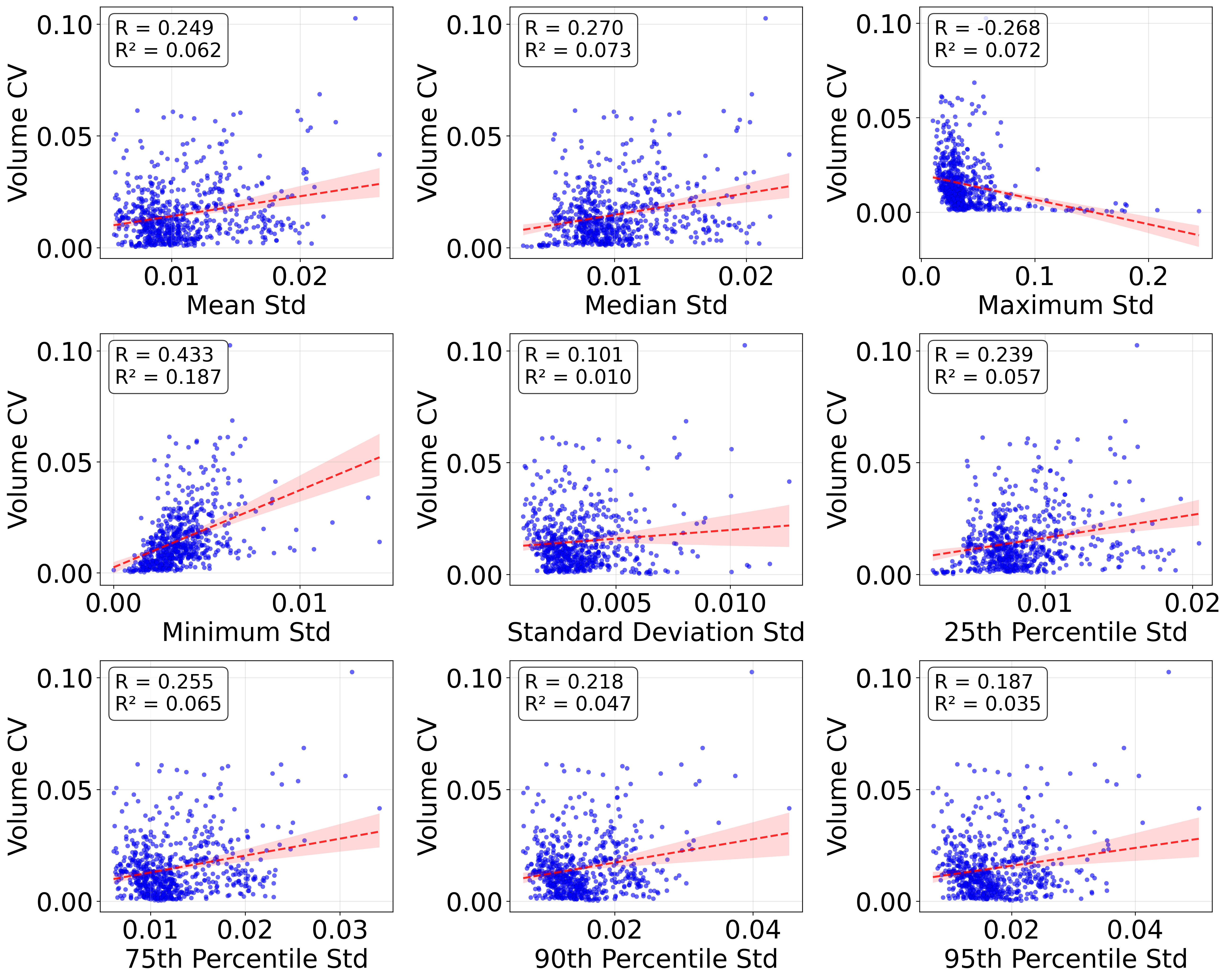}
    \caption{Scatter plots show correlations between nine summary statistics of voxelwise standard deviation and volume coefficient of variation (CV) for each brain structure. Each point represents one structure from one validation subject. Volume CV removes structure size as a confounding factor by normalizing standard deviation by mean volume. Low correlations indicate that voxelwise uncertainty maps poorly predict clinically relevant volume measurement uncertainty, supporting the need for task-specific uncertainty evaluation methods in accelerated MRI assessment.}
    \label{fig:VN_correlation}
\end{figure}

We next tested whether voxelwise intensity uncertainty could explain volume variability. For each structure, we computed voxel-level standard deviation maps from the ensemble, summarizing them with statistics (mean, median, max, min, std, selected percentiles) within the union of ensemble segmentations. We correlated these voxelwise uncertainty summaries with normalized volume standard deviation to evaluate predictive relationships. Figure \ref{fig:VN_correlation} shows representative results for the Variational Network (VN); results for other architectures are provided in the appendix. Correlations between voxel-level uncertainty statistics and structure-level volume variability were consistently low, suggesting no meaningful relationship.  

Additionally, we tested whether volume standard deviation could be predicted from normalized histograms of voxelwise standard deviation (binned into 10 bins) and from summary statistics of these histograms in separate regression analyses. The regression analyses yielded low $R^2$ values, indicating poor predictive power: \textbf{UNET (0.1766, 0.2482)},\; \textbf{VN (0.1352, 0.1605)},\; \textbf{RIM (0.2425, 0.3204)}. These results demonstrate that voxel-level uncertainty does not reliably predict ensemble-level volume variability, underscoring the need for task-specific evaluation of uncertainty in accelerated MRI reconstruction.

\section{Discussion}
Uncertainty quantification is essential for the clinical adoption of  DL-based accelerated MRI reconstruction. Our results show that assessing uncertainty in isolation, without explicitly tying it to the intended clinical task, can fail to reveal critical errors and biases introduced during reconstruction. We demonstrated that undersampled reconstructions can produce systematic morphological variability and segmentation bias relative to fully sampled references. These effects are not adequately captured by voxel-level or global uncertainty metrics alone. Critically, such variability can directly impact clinical assessments of neurodegenerative diseases like Alzheimer's or Huntington's \cite{freeborough_boundary_1997}. To ensure safe deployment, uncertainty quantification in reconstruction algorithms must be task-specific, directly reflecting the reliability of downstream analyses.

Further reinforcing this point, our correlation and regression analyses revealed that ensemble-level variability in volume estimates is poorly predicted by common uncertainty metrics such as voxelwise intensity variation. This disconnect highlights that existing measures may be fundamentally inadequate when segmentation and volumetry are the clinical targets. Taken together, our findings argue for a more structure-aware, task-specific approach to evaluating reconstruction uncertainty. We believe it is necessary to explicitly quantify bias and error in morphological measurements aligned with real clinical objectives.

Lastly, we note that all models investigated in this study exhibit high image quality when assessed using standard metrics such as SSIM and PSNR. However, PSNR is purely pixelwise and ignores structural correlation, while SSIM is structure-aware but not task-aware. Comparison of the results in Figure \ref{fig:Distribution of CV values across samples} with SSIM and PSNR scores indicates that within-ensemble variability is not accurately captured by either metric. For example, although the VN and RIM models achieve similar PSNR and SSIM values, they show substantial differences in morphological uncertainty. Conversely, the UNet and VN differ more in SSIM and PSNR scores, yet have comparable volume uncertainty. This raises a critical question about whether these image quality metrics should remain the de facto standard for evaluating reconstruction performance. While they offer convenient, standardized measures of fidelity, they may fail to detect clinically meaningful biases or variations arising from accelerated acquisition and learned reconstruction. Overall, our results suggest the need for a more nuanced evaluation approach that explicitly considers downstream clinical tasks and the interpretability of uncertainty, rather than relying solely on pixel-level similarity.

\subsection{Further Research}
While the results presented here provide evidence against intensity-based uncertainty quantification, we acknowledge several limitations. Relying on a single segmentation architecture is a weakness, and future research should incorporate different segmentation models. We used SynthSeg \cite{billot_robust_2023} as a representative downstream task because of its accessibility and ease of use. However, its deterministic nature prevents formal disentanglement of segmentation-induced from reconstruction-induced uncertainty. Despite this, observed variability in segmentation outcomes still serves as a valid measure of pipeline-level uncertainty. Future work would benefit from employing probabilistic segmentation models to better separate uncertainty originating from the segmentation process itself versus that introduced by reconstruction.  Additionally, while segmentation is an important clinical target, other tasks such as cortical thickness estimation may offer further insight into the consequences of reconstruction uncertainty. Finally, a persistent limitation is the lack of high-quality pathological k-space data. The samples analyzed here are from healthy subjects, whereas MRI is typically used for diagnosing conditions such as neurodegenerative diseases. In these contexts, it is critical for reconstruction algorithms to exhibit predictable epistemic uncertainty. Given that generative models can remove structures not seen during training \cite{cohen_distribution_2018}, it is essential to assess task-specific uncertainty quantification on pathological data.

\section{Conclusion}
In this work, we evaluated existing approaches to uncertainty quantification in learned MRI reconstruction algorithms. We found that variability and bias introduced by machine learning methods in accelerated MRI reconstruction are not adequately captured by common image quality or intensity-based uncertainty metrics. Our results highlight the need for uncertainty quantification in reconstruction tasks to move beyond intensity-based measures. In conclusion, we argue that uncertainty in accelerated MRI reconstruction should be evaluated in the context of clinical tasks rather than solely in pixel space.

\newpage
\bibliographystyle{splncs04}
\bibliography{references.bib}

\section{Appendix}
\begin{table}
\begin{center}
\caption{Overview of the architectures used in this study, with ensemble size, batch size, training steps, and average SSIM and PSNR.}\label{tab1}
\begin{tabular}{|l|l|l|l|l|l|}
\hline
Model & Ensemble & Batch Size & Steps & SSIM (avg.) & PSNR (avg.) \\
\hline
UNet & 15 & 8 & 100000 & 0.8877 $\pm$ 0.0024 & 30.65 $\pm$ 0.18 \\
VarNet & 12 & 8 & 100000 & 0.9164 $\pm$ 0.0127 & 33.40 $\pm$ 1.09 \\
RIM & 11 & 2 & 100000 & 0.9329 $\pm$ 0.0003 & 35.05 $\pm$ 0.04 \\ 
\hline
\end{tabular}
\end{center}
\end{table}

\begin{figure}
    \centering
    \includegraphics[width=0.8\linewidth]{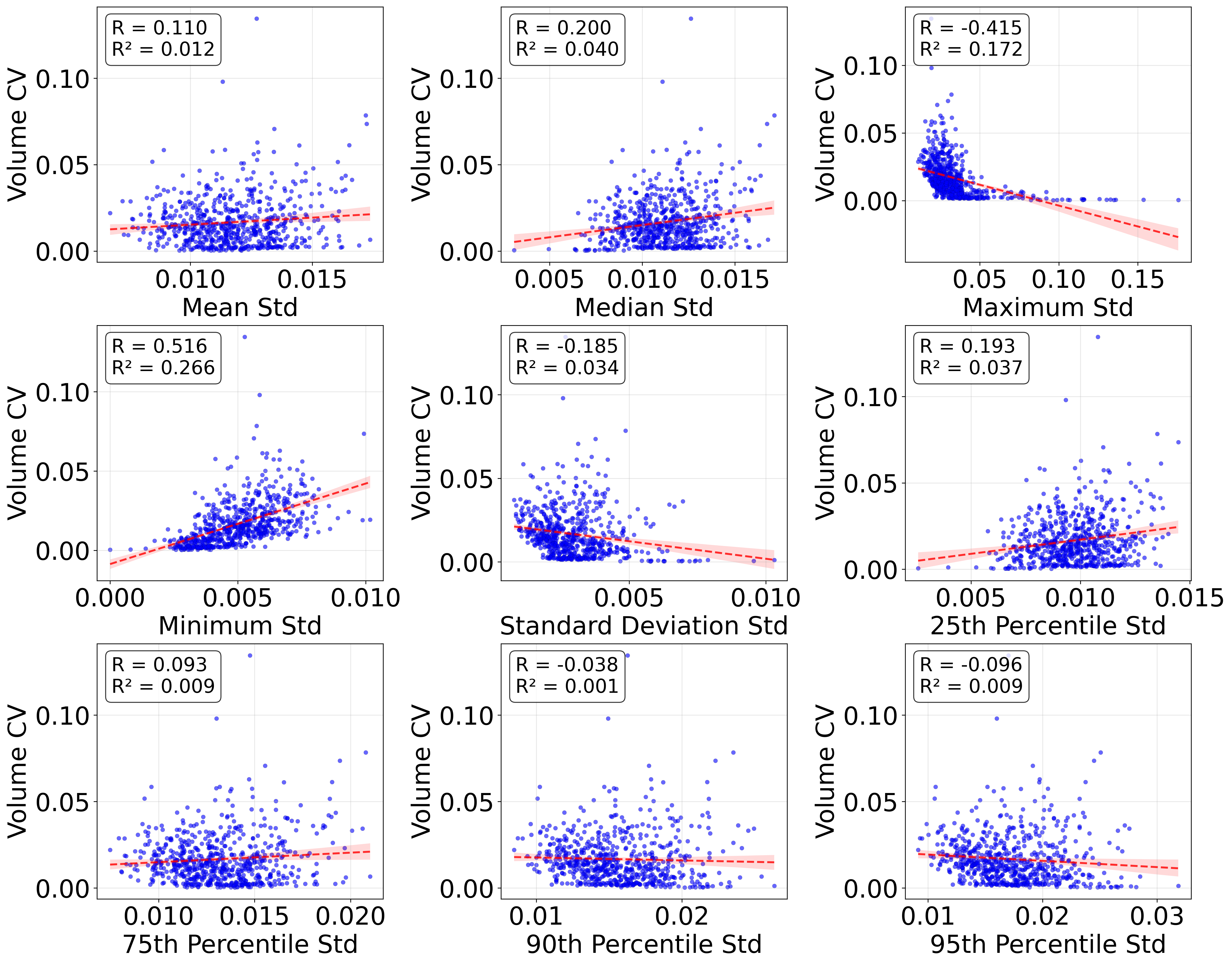}
    \caption{Scatter plots show correlations between nine summary statistics of voxelwise standard deviation and volume coefficient of variation (CV) for each brain structure for the UNET.}
    \label{fig:ecorrelation_plots_UNET}
\end{figure}

\begin{figure}
    \centering
    \includegraphics[width=0.8\linewidth]{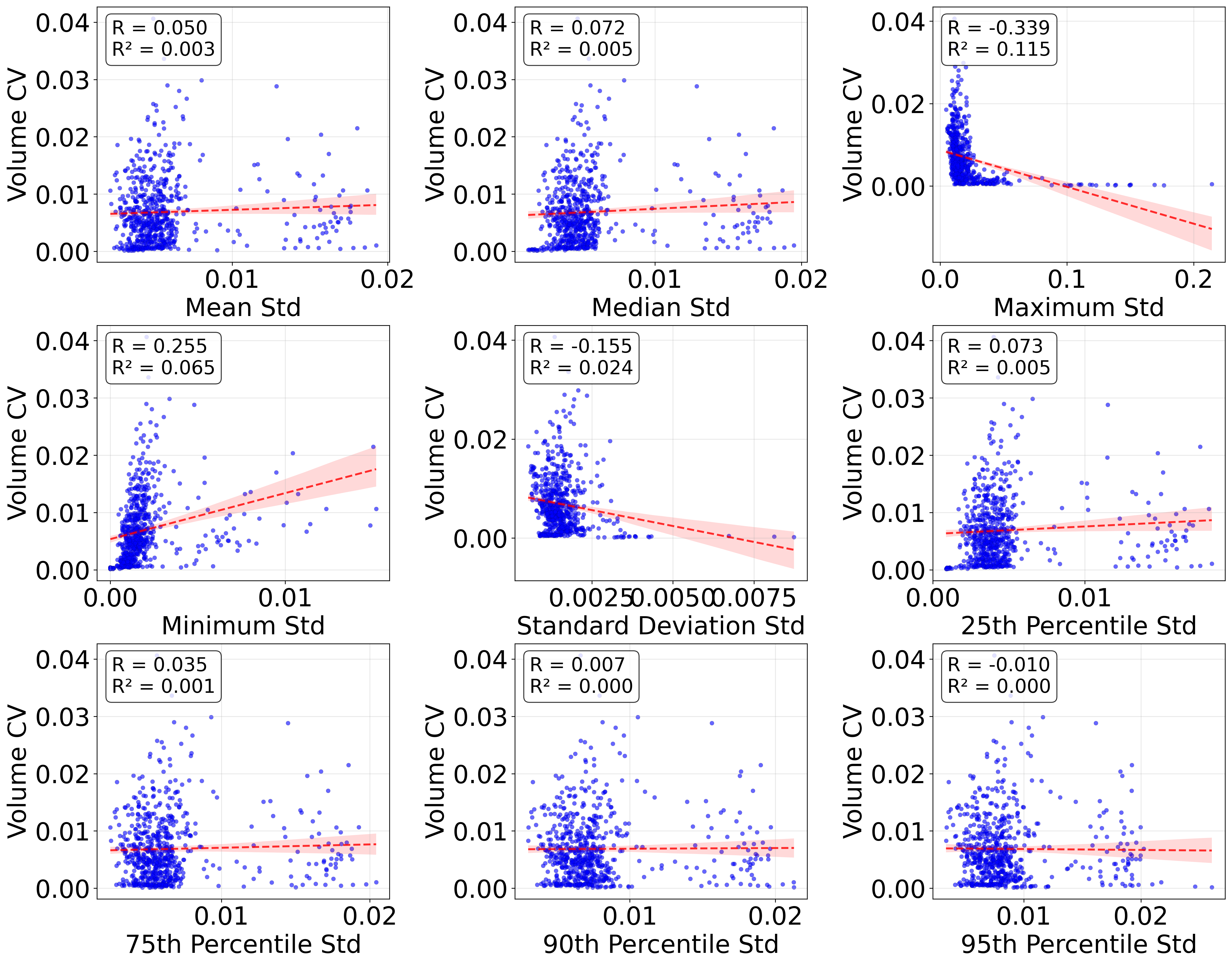}
    \caption{Scatter plots show correlations between nine summary statistics of voxelwise standard deviation and volume coefficient of variation (CV) for each brain structure for the RIM.}
    \label{fig:ecorrelation_plots_RIM}
\end{figure}
\end{document}